\documentclass[preprint,preprintnumbers,amsmath,amssymb]{revtex4}
\pdfoutput=1
\usepackage{epsfig}
\usepackage{graphicx}
\usepackage{dcolumn}
\usepackage{bm}
\usepackage{threeparttable}
\usepackage{xcolor}
\usepackage[utf8]{inputenc}
\usepackage{booktabs}

\def\beq{\begin{equation}}
\def\eeq{\end{equation}}

\begin{document}

\title{ Exploring multi-step electroweak phase transitions in the 2HDM+$\bm{a}$ }

\author{Zong-guo Si$^{1}$}
\author{Hong-xin Wang$^{1}$}
\author{Lei Wang$^{2}$}
\author{Yang Zhang$^{3,4}$}

\affiliation{$^1$School of Physics, Shandong University, Jinan, Shandong 250100, China\\
$^2$Department of Physics, Yantai University, Yantai, Shandong 264005, China\\
$^3$School of Physics, Zhengzhou University, Zhengzhou 450000, China\\
$^4$School of Physics, Henan Normal University, Xinxiang 453007, China
}


\begin{abstract}
Multiple electroweak phase transitions occurring sequentially in the early universe can give rise to intriguing phenomenology, compared to the typical single-step electroweak phase transition. In this work, we investigate this scenario within the framework of the two-Higgs-doublet model with a pseudoscalar, utilizing the complete one-loop finite-temperature effective potential.
After considering relevant experimental and theoretical constraints, we identify four distinct types of phase transitions.
In the first case, only the configuration of the CP-even Higgs acquires a non-zero value via a first-order or a cross-over electroweak phase transition, leading to electroweak symmetry breaking. In the remaining three cases, the pseudoscalar fields can obtain vacuum expectation values at different phases of the multi-step phase transition process, leading to the spontaneous breaking of the CP symmetry. As the temperature decreases, the phase shifts to the vacuum observed today via first-order electroweak phase transition, at this point, the vacuum expectation value of the pseudoscalar field returns to zero, restoring the CP symmetry.
Finally, we compare the transition strength and the stochastic gravitational wave background generated in the four situations along with the projected detection limits.

\end{abstract}

\maketitle

\section{Introduction}
In the intersecting of particle physics and cosmology, exploring a natural explanation for baryon asymmetry of the universe (BAU) \cite{matter_antimatter} remains a challenging topic.
Three Sakharov conditions must be satisfied for a dynamical generation of BAU:
 baryon number violation, C and CP violation, and departure from thermal equilibrium \cite{sakharov}. 
Among the mechanisms  \cite{GUT_1,GUT_2,Affleck-Dine,EWBG1,EWBG2,Leptogenesis}, the electroweak baryogenesis is an attractive mechanism for explaining the BAU due to its testability at the LHC and at the precision frontier by the electric dipole moment (EDM) experiments. 
The condition of departure from thermal equilibrium is met via a strong first-order electroweak phase transition (PT) \cite{SFOEWPT_1,SFOEWPT_2}. 
However, the electroweak PT in the SM is a crossover \cite{80GeV_1,80GeV_2}, and the CP violation
provided by the CKM matrix is too small to explain the observed BAU \cite{SM_CPV}.
Therefore, the SM needs to be extended to provide large CP violation and a strong first-order electroweak PT, such as the singlet extension of SM (see e.g. \cite{bgs-1,bgs-2,bgs-3,bgs-4,bgs-5,bgs-6,bgs-7,Huang:2018aja,Xie:2020wzn,cao,huang,Xiao:2022oaq,Xiao:2023dbb,Balazs:2023kuk,Wang:2023suf}), the
two-Higgs-doublet model (2HDM) and its extensions (see e.g. \cite{bg2h-1,bg2h-2,bg2h-3,bg2h-4,bg2h-5,bg2h-6,bg2h-7,bg2h-8,bg2h-9,Goncalves:2016iyg,bg2h-10,bg2h-11,bg2h-12,Athron:2019teq,bg2h-13,Basler:2020nrq,Han:2020ekm,2111.13079,Goncalves:2021egx,Wang:2022yhm,2207.00060,Arcadi:2022lpp}).
Adding explicit CP violation to these models can affect the EDM, which is constrained by negative results from electron EDM searches \cite{EDM}. 
A finite temperature spontaneous CP violation mechanism can produce sufficient large CP violation while satisfying the EDM data naturally. 
Here, the CP symmetry is spontaneously broken at high temperature and restored in the current universe. The interesting mechanism has been implemented in the singlet scalar extension of
the SM \cite{cao,huang}, the singlet pseudoscalar extension of 2HDM (2HDM+$\bm{a}$) \cite{2hdm+s_2,Liu:2023sey} and the complex singlet scalar extension of 2HDM \cite{Ma:2023kai,Gao:2024qag}.

In the 2HDM+$\bm{a}$, mixing between the singlet pseudoscalar and the Higgs doublet pseudoscalar occurs at zero temperature.   
At high temperature, these pseudoscalar fields acquire nonzero vacuum expectation value~(VEV) leading to spontaneous CP symmetry breaking. 
This process can generate the observed BAU  during a strong first-order electroweak PT.
Ultimately, the observed vacuum state emerges, restoring CP symmetry at present temperatures. 
In Ref.~\cite{2hdm+s_2,Liu:2023sey},
the high-temperature approximation of the effective potential is taken to analyze the PT, which includes only the tree-level scalar potential 
and thermal masses of background fields.

In this work, we analyze the PTs in 2HDM+$\bm{a}$ using the full one-loop finite-temperature effective potential which includes the tree-level potential, the Coleman-Weinberg term \cite{Vcw1,Vcw2} and the finite-temperature corrections \cite{vth}. 
We employ the $\overline{\rm MS}$ scheme to handle the one-loop corrected tadpole conditions and Higgs masses, avoiding the infrared divergences associated with Goldstone boson loops that arise in the on-shell scheme. 
The model exhibits various PT patterns, including several multi-step PTs that deserve further detailed study. 
A first-order PT can generate detectable stochastic gravitational wave (GW) signals, which provide a new approach to search for new physics~\cite{Athron:2023xlk}.  
Compared to single-step PT, multi-step PTs can generate more diverse ranges of GW signals, with broader features in peak amplitude and frequency.

This paper is organized as follows: 
In Section \ref{sec2}, we provide a review of the 2HDM+$\bm{a}$ and present the full one-loop finite-temperature effective potential. 
In Section \ref{sec3}, we introduce theoretical and experimental constraints on the parameter space of the model. 
In Section \ref{sec4}, we discuss four different types of PTs and illustrate the evolution of the universe.
In Section \ref{sec5}, we numerically assess the potential of future GW detectors to probe the multi-step PTs identified in our work.
Finally, we present our conclusion in Section \ref{sec6}.

\section{2HDM+$\bm{a}$ and the finite temperature effective potential }\label{sec2}
We extend the 2HDM by a pseudoscalar singlet field, whose tree-level potential is written as
\begin{equation}
	\begin{aligned}
		\label{Vtree}
		{V_0}(\Phi_1,\Phi_2,S) &= 
		m_{11}^2\left( {\Phi _1^\dag {\Phi _1}} \right) + m_{22}^2\left( {\Phi _2^\dag {\Phi _2}} \right) - \left[ {m_{12}^2\Phi _1^\dag {\Phi _2} + {\rm{ h}}{\rm{.c}}{\rm{. }}} \right]\\
		{\rm{             }} 
		&+ \frac{\lambda _1}{2}{\left( {\Phi _1^\dag {\Phi _1}} \right)^2} + \frac{\lambda _2}{2}{\left( {\Phi _2^\dag {\Phi _2}} \right)^2} + {\lambda _3}\left( {\Phi _1^\dag {\Phi _1}} \right)\left( {\Phi _2^\dag {\Phi _2}} \right)\\
		{\rm{             }} 
		&+ {\lambda _4}\left( {\Phi _1^\dag {\Phi _2}} \right)\left( {\Phi _2^\dag {\Phi _1}} \right) + \left[ {\frac{\lambda _5}{2}{{\left( {\Phi _1^\dag {\Phi _2}} \right)}^2} + {\rm{ h}}{\rm{.c}}{\rm{. }}} \right]\\
		{\rm{             }}
		&+ \frac{1}{2}m_0^2{S^2} + \frac{{{\kappa _S}}}{{24}}{S^4} + \left[ {i\mu S\Phi _2^\dag {\Phi _1} + {\rm{ h}}{\rm{.c}}{\rm{. }}} \right] + \frac{{{\kappa _1}}}{2}{S^2}\Phi _1^\dag {\Phi _1} + \frac{{{\kappa _2}}}{2}{S^2}\Phi _2^\dag {\Phi _2}.
	\end{aligned}
\end{equation}
To maintain the CP conservation, we assume all the mass and coupling parameters are real, and the singlet field $S$ has no VEV at zero temperature.
The $\Phi_1$ and $\Phi_2$ are the two Higgs-doublet fields,
\begin{equation}
	\Phi_1=\left(\begin{array}{c} \phi_1^+ \\
		\frac{1}{\sqrt{2}}\,(v_1+\phi_1+i\eta_1)
	\end{array}\right)\,, \ \ \
	\Phi_2=\left(\begin{array}{c} \phi_2^+ \\
		\frac{1}{\sqrt{2}}\,(v_2+\phi_2+i\eta_2)
	\end{array}\right)\,
\end{equation}
where $v_1$ and $v_2$ are VEVs with $v^2 = v^2_1 + v^2_2 = (246~\rm GeV)^2$, and the ratio of the two VEVs is defined as
\begin{equation}
	\tan \beta  \equiv \frac{{{v_2}}}{{{v_1}}}.
\end{equation}

The general Yukawa interactions at tree-level order are written as
\begin{equation}\label{eq8}
	\begin{aligned}
		- {\cal L} = 
		&{Y_{u1}}{{\bar Q}_L}{{\tilde \Phi }_1}{u_R} + {Y_{u2}}{{\bar Q}_L}{{\tilde \Phi }_2}{u_R}{\rm{ }}\\
		&+ {Y_{d1}}{{\bar Q}_L}{\Phi _1}{d_R} + {Y_{d2}}{{\bar Q}_L}{\Phi _2}{d_R}\\
		&+ {Y_{l1}}{{\bar L}_L}{\Phi _1}{e_R} + {Y_{l2}}{{\bar L}_L}{\Phi _2}{e_R} + h.c.,
	\end{aligned}
\end{equation}
where $Q_L^T=({{u_L},{\rm{ }}{d_L}}),{\rm{ }}L_L^T = ( {{\nu _L},{\rm{ }}{l_L}} ),{\rm{ }}\tilde \Phi _{k} = i{\tau _2}\Phi _{k}^*$ with $k=1,2$. The elements of ${Y_{uk}},{\rm{ }}{Y_{dk}}$ and ${\rm{ }}{Y_{lk}}$ determine the interactions between different scalar fields and fermions. We assume the Yukawa interactions to be aligned in order to prevent the tree-level flavor changing neutral current \cite{FCNC,aligned}, 
\begin{equation}
	\begin{aligned}
		&{\left( {{Y_{u1}}} \right)_{ii}} = \frac{{\sqrt 2 {m_{ui}}}}{v}\left( {{c_\beta } - {s_\beta }{\kappa _u}} \right),{\left( {{Y_{u2}}} \right)_{ii}} = \frac{{\sqrt 2 {m_{ui}}}}{v}\left( {{s_\beta } + {c_\beta }{\kappa _u}} \right),\\
		&{\left( {{Y_{l1}}} \right)_{ii}} \;= \frac{{\sqrt 2 {m_{li}}}}{v}\left( {{c_\beta } - {s_\beta }{\kappa _l}} \right),\;\;{\left( {{Y_{l2}}} \right)_{ii}} = \frac{{\sqrt 2 {m_{li}}}}{v}\left( {{s_\beta } + {c_\beta }{\kappa _l}} \right),\\
		&{\left( {{X_{d1}}} \right)_{ii}} = \frac{{\sqrt 2 {m_{di}}}}{v}\left( {{c_\beta } - {s_\beta }{\kappa _d}} \right),{\left( {{X_{d2}}} \right)_{ii}} = \frac{{\sqrt 2 {m_{di}}}}{v}\left( {{s_\beta } + {c_\beta }{\kappa _d}} \right).
	\end{aligned}
\end{equation}
$i=1,2,3$ is the index of generation and ${X_{d1,2}} = V_{dL}^\dagger Y_{d1,2} V_{dR}$ with $V_{dL}\equiv V_{CKM}$.  The unitary matrices $V_{dL,R}$ transform the original  eigenstates to the mass
eigenstates for the left-handed and right-handed down-type quark fields. The type-I (type-II) 2HDM can be realised by choices of 
$\kappa_u=\kappa_d=\kappa_\ell=1/\tan\beta$  ($\kappa_u=1/\tan\beta$ and $\kappa_d=\kappa_\ell=-\tan\beta$).

We analyze the phase histories of the model in the classical field space spanned by
\begin{equation}
	\langle\Phi_1 \rangle=\left(\begin{array}{c} 0 \\
		\frac{1}{\sqrt{2}}\,h_1
	\end{array}\right)\,, \ \ \
	\langle\Phi_2 \rangle=\left(\begin{array}{c} 0 \\
		\frac{1}{\sqrt{2}}\,(h_2+ih_3)
	\end{array}\right)\,, \ \ \
	\langle S \rangle=h_4.
\end{equation}
Here we take the imaginary part of the neutral elements of $\langle\Phi_1 \rangle$ to be zero.
This choice is reasonable since the potential of Eq.~(\ref{Vtree}) only depends on
the relative phase of the neutral elements of $\langle\Phi_1 \rangle$ and $\langle\Phi_2 \rangle$.

The Higgs potential is modified from its tree-level form by the radiative corrections. At zero temperature, the effective potential at the one-loop level is written as,
 \begin{equation}
V_{\rm{eff,T=0}}({{h_1},{h_2},{h_3},{h_4}}) =  V_{0}({{h_1},{h_2},{h_3},{h_4}}) +  V_{\rm{CW}}({{h_1},{h_2},{h_3},{h_4}}) 
\label{one-loop effective potential}
 \end{equation}
where $V_{0}$ is the tree-level potential, and $V_{\rm{CW}}$ is the Coleman-Weinberg (CW) potential.
The $V_{0}$ is written as,
\begin{equation}
	\begin{aligned}
		{V_0}\left( {{h_1},{h_2},{h_3},{h_4}} \right) =
		&\frac{1}{2}m_{11}^2h_1^2 + \frac{1}{2}m_{22}^2h_2^2 + \frac{1}{2}m_{22}^2h_3^2 - m_{12}^2{h_1}{h_2}\\
		{\rm{                          }} 
		&+ \frac{1}{8}{\lambda _1}h_1^4 + \frac{1}{8}{\lambda _2}h_2^4 + \frac{1}{8}{\lambda _2}h_3^4 + \frac{1}{4}h_2^2h_3^2{\lambda _2} + \frac{1}{4}h_1^2h_2^2{\lambda _{345}} + \frac{1}{4}h_1^2h_3^2{k_{345}}\\
		{\rm{                          }} 
		&+ \mu{h_1}{h_3}{h_4} + \frac{1}{2}m_0^2h_4^2  + \frac{{{\kappa _S}}}{{24}}h_4^4 + \frac{{{\kappa _1}}}{4}h_4^2h_1^2 + \frac{{{\kappa _2}}}{4}h_4^2\left( {h_2^2 + h_3^2} \right),
	\end{aligned}
\end{equation}
where $k_{345}=\lambda_3+\lambda_4-\lambda_5$ and ${\lambda _{345}} = {\lambda _3} + {\lambda _4} + {\lambda _5}$. 

The $V_{\rm{CW}}$ in the $\overline{\rm MS}$  renormalization scheme is given by  \cite{Vcw2}
\begin{equation}
	{V_{{\rm{CW}}}}\left( {{h_1},{h_2},{h_3},{h_4}} \right) = \sum\limits_i {{{( - 1)}^{2{s_i}}}} {n_i}\frac{{\hat m_i^4\left( {{h_1},{h_2},{h_3},{h_4}} \right)}}{{64{\pi ^2}}}\left[ {\ln \frac{{\hat m_i^2\left( {{h_1},{h_2},{h_3},{h_4}} \right)}}{{{Q^2}}} - {C_i}} \right],
	\label{Vcw}
\end{equation}
 $Q$ is a renormalization scale, and we take $Q^2=v^2$.
$\hat{m_i}$ is the background-field-dependent mass
of the neutral scalars $\left( {h, H, A, \bm{a}, G} \right)$, the charged scalars $\left( {{H^ \pm },{G^ \pm }} \right)$, 
the heavy SM quark $\left( {t} \right)$ and the gauge bosons $\left( {Z,{W^ \pm }} \right)$.
The expression of $\hat{m_i}$ can be found in the Appendix A.
$s_i$ is the spin of particle $i$, and 
the constant $C_i =\frac{3}{2}$ for scalars or fermions and
$C_i = \frac{5}{6}$ for gauge bosons.
$n_i$ is the number of degrees of freedom,
\begin{align}
&n_h=n_H=n_A=n_{\bm{a}}=n_G=1,\nonumber\\ 
&n_{H^\pm}=n_{G^\pm}=2,\nonumber\\
&n_{W^\pm}=6,~n_{Z}=3,\nonumber\\
&n_{t}=12.
\end{align}

We choose the $\overline{\rm MS}$ renormalization scheme and the Landau gauge for the one-loop zero-temperature effective potential Eq.~(\ref{one-loop effective potential}).  The one-loop corrected tadpole conditions,
\begin{equation}
	\begin{aligned}
		&\left\langle{ {\frac{{\partial ({V_0} + {V_{\rm{CW}}}})}{{\partial {h_1}}}} }\right\rangle = 0{\rm{  ,    }}\\
		&\left\langle{ {\frac{{\partial ({V_0} + {V_{\rm{CW}}}})}{{\partial {h_2}}}} }\right\rangle = 0{\rm{ .}}
	\end{aligned}
 \label{renormalization-1}
\end{equation}
The angle bracket $\left\langle \cdots \right\rangle$ denotes the corresponding field-dependent values
evaluated at the chosen true vacuum ($h_1=v_1$, $h_2=v_2$, $h_3=0$, and $h_4=0$) at zero temperature.

The one-loop improved mass squared matrices of the CP-even scalars, the charged scalars, and the CP-odd scalars are
\begin{equation}
	\left\langle {\begin{array}{*{20}{c}}
			{\frac{{{\partial ^2}\left( {{V_0} + {V_{{\rm{CW}}}}} \right)}}{{\partial h_1^2}}}&{\frac{{{\partial ^2}\left( {{V_0} + {V_{{\rm{CW}}}}} \right)}}{{\partial {h_1}\partial {h_2}}}}\\
			{\frac{{{\partial ^2}\left( {{V_0} + {V_{{\rm{CW}}}}} \right)}}{{\partial {h_2}\partial {h_1}}}}&{\frac{{{\partial ^2}\left( {{V_0} + {V_{{\rm{CW}}}}} \right)}}{{\partial h_2^2}}}
	\end{array}} \right\rangle = \left( {\begin{array}{*{20}{c}}
			{{c_\alpha }}&{ - {s_\alpha }}\\
			{{s_\alpha }}&{{c_\alpha }}
	\end{array}} \right)\left( {\begin{array}{*{20}{c}}
			{{m^2_H}}&0\\
			0&{{m^2_h}}
	\end{array}} \right)\left( {\begin{array}{*{20}{c}}
			{{c_\alpha }}&{{s_\alpha }}\\
			{ - {s_\alpha }}&{{c_\alpha }}
	\end{array}} \right),
\end{equation}
\label{renormalization-2}
\begin{equation}
\left\langle {\begin{array}{*{20}{c}}
		{\frac{{{\partial ^2}\left( {{V_0} + {V_{{\rm{CW}}}}} \right)}}{{\partial {G^ + }\partial {G^ - }}}}&{\frac{{{\partial ^2}\left( {{V_0} + {V_{{\rm{CW}}}}} \right)}}{{\partial {G^ + }\partial {H^ - }}}}\\
		{\frac{{{\partial ^2}\left( {{V_0} + {V_{{\rm{CW}}}}} \right)}}{{\partial {H^ + }\partial {G^ - }}}}&{\frac{{{\partial ^2}\left( {{V_0} + {V_{{\rm{CW}}}}} \right)}}{{\partial {H^ + }\partial {H^ - }}}}
\end{array}} \right\rangle  = \left( {\begin{array}{*{20}{c}}
		{{c_\beta }}&{{s_\beta }}\\
		{ - {s_\beta }}&{{c_\beta }}
\end{array}} \right)\left( {\begin{array}{*{20}{c}}
		{{m^2_{{G^ \pm }}}}&0\\
		0&{{m^2_{{H^ \pm }}}}
\end{array}} \right)\left( {\begin{array}{*{20}{c}}
		{{c_\beta }}&{ - {s_\beta }}\\
		{{s_\beta }}&{{c_\beta }}
\end{array}} \right),
\label{renormalization-3}
\end{equation}
\begin{equation}
	\left\langle {\begin{array}{*{20}{c}}
			{\frac{{{\partial ^2}\left( {{V_0} + {V_{{\rm{CW}}}}} \right)}}{{\partial {G^2}}}}&{\frac{{{\partial ^2}\left( {{V_0} + {V_{{\rm{CW}}}}} \right)}}{{\partial G\partial {h_3}}}}&{\frac{{{\partial ^2}\left( {{V_0} + {V_{{\rm{CW}}}}} \right)}}{{\partial G\partial {h_4}}}}\\
			{\frac{{{\partial ^2}\left( {{V_0} + {V_{{\rm{CW}}}}} \right)}}{{\partial {h_3}\partial G}}}&{\frac{{{\partial ^2}\left( {{V_0} + {V_{{\rm{CW}}}}} \right)}}{{\partial h_3^2}}}&{\frac{{{\partial ^2}\left( {{V_0} + {V_{{\rm{CW}}}}} \right)}}{{\partial {h_3}\partial {h_4}}}}\\
			{\frac{{{\partial ^2}\left( {{V_0} + {V_{{\rm{CW}}}}} \right)}}{{\partial {h_4}\partial G}}}&{\frac{{{\partial ^2}\left( {{V_0} + {V_{{\rm{CW}}}}} \right)}}{{\partial {h_4}\partial {h_3}}}}&{\frac{{{\partial ^2}\left( {{V_0} + {V_{{\rm{CW}}}}} \right)}}{{\partial h_4^2}}}
	\end{array}} \right\rangle  = U\left( {\begin{array}{*{20}{c}}
			{{m^2_G}}&0&0\\
			0&{{m^2_A}}&0\\
			0&0&{{m^2_{\bm{a}}}}
	\end{array}} \right)U^T,
 \label{renormalization-4}
\end{equation}
with 
\begin{equation}
	U= \left( {\begin{array}{*{20}{c}}
			{{c_\beta }}&{{s_\beta }}&0\\
			{ - {c_\theta }{s_\beta }}&{{c_\beta }{c_\theta }}&{-{s_\theta }}\\
			{-{s_\beta }{s_\theta }}&{{c_\beta }{s_\theta }}&{{c_\theta }}
	\end{array}} \right).
\end{equation}
After solving the renormalization equations Eq.~(\ref{renormalization-1}) to Eq.~(\ref{renormalization-4}), the potential parameters in Eq.~(\ref{Vtree}) are determined, and the background-field-dependent masses no longer satisfying tree-level relations, particularly ${m^2_{{G},{G^ \pm }}}$$\neq$ 0.
Therefore, the second derivative of $V_{\rm{CW}}$ is free from infrared divergence \cite{IR,2208.01319}, avoiding the infrared divergences
associated with Goldstone boson loops that arise in the on-shell scheme.

The physical scalar states are two neutral CP-even states $h$ and $H$, 
 two neutral pseudoscalars $A$ and $\bm{a}$, and a pair of charged scalars $H^{\pm}$. 
Their Yukawa couplings are given by
\begin{equation}
	\begin{aligned}
		- {{\cal L}_Y} &= \frac{{{m_f}}}{v}(\sin \left( {\beta  - \alpha } \right) + \cos \left( {\beta  - \alpha } \right){\kappa _f})h\overline f f \\
		&+ \frac{{{m_f}}}{v}(\cos \left( {\beta  - \alpha } \right) - \sin \left( {\beta  - \alpha } \right){\kappa _f})H\overline f f\\
		{\rm{          }} 
		&- i\frac{{{m_u}}}{v}{\kappa _u}{c_\theta }A\bar u{\gamma _5}u + i\frac{{{m_d}}}{v}{\kappa _d}{c_\theta }A\overline d {\gamma _5}d + i\frac{{{m_l}}}{v}{\kappa _l}{c_\theta }A\overline l {\gamma _5}l\\
		{\rm{          }} 
		&- i\frac{{{m_u}}}{v}{\kappa _u}{s_\theta }\bm{a}\bar u{\gamma _5}u + i\frac{{{m_d}}}{v}{\kappa _d}{s_\theta }\bm{a}\overline d {\gamma _5}d + i\frac{{{m_l}}}{v}{\kappa _l}{s_\theta }\bm{a}\overline l {\gamma _5}l\\
		{\rm{          }} 
		&+ {H^ + }\bar u{V_{CKM}}\left( {\frac{{\sqrt 2 {m_d}}}{v}{\kappa _d}{P_R} - \frac{{\sqrt 2 {m_u}}}{v}{\kappa _u}{P_L}} \right)d + {\rm{ h}}{\rm{.c}}{\rm{. }}\\
		{\rm{          }} 
		&+ \frac{{\sqrt 2 {m_l}}}{v}{\kappa _l}{H^ + }\bar v{P_R}e + {\rm{ h}}{\rm{.c}}..{\rm{ }}
	\end{aligned}
\end{equation}
 
 At finite temperatures, the thermal corrections to the potential become important. Here we use the one-loop correction \cite{vth},
 \begin{equation}
 	\begin{aligned}
 		&{V_{{\rm{th}}}}\left( {{h_1},{h_2},{h_3},{h_4},T} \right) = \frac{{{T^4}}}{{2{\pi ^2}}}\sum\limits_i {{n_i}} {J_{B,F}}\left( {\frac{{\hat m_i^2\left( {{h_1},{h_2},{h_3},{h_4}} \right)}}{{{T^2}}}} \right),\\
 		&{J_{B,F}}(y) =  \pm \int_0^\infty  d x{x^2}\ln \left[ {1 \mp \exp \left( { - \sqrt {{x^2} + y} } \right)} \right].
 	\end{aligned}
 \end{equation}
In this work, we employ the Parwani method~\cite{Parwani} to handle the daisy diagrams~\cite{A-E}. This involves replacing the mass $\hat m_i^2{{\left( {{h_1},{h_2},{h_3},{h_4}} \right)}}$  with the thermal Debye mass $\bar M_i^2{{\left( {{h_1},{h_2},{h_3},{h_4},T} \right)}}$. 
The thermal Debye masses for $h, H, A, \bm{a}, G, H^\pm, G^\pm, W^\pm_L, Z_L, \gamma_L$ can be found in Appendix A, where $W^\pm_L,~Z_L$, and $\gamma_L$ are the longitudinal gauge bosons with $n_{W^\pm_L}=2,~n_{Z_L}=n_{\gamma_L}=1$. 
\section{THEORETICAL AND EXPERIMENTAL CONSTRAINTS}\label{sec3}
{\bf (1) Theoretical constraints.}
The requirement of vacuum stability ensures that the potential is bounded from
below and that the electroweak vacuum is the global minimum of the full scalar potential.

The unitarity constraint is applied to prevent theoretical $2 \to 2$ particle scattering processes from producing unphysical results. Detailed discussions on the vacuum stability and unitarity can be found in Ref.~\cite{1408.2106,1609.03551}, we employ the formulas in \cite{1408.2106,1609.03551} to implement the relevant theoretical constraints.

{\bf (2) The signal data of the 125 GeV Higgs.}
 We take the light CP-even $h$ as the observed 125 GeV Higgs, and $\sin(\beta-\alpha) \approx 1$ to be aligned to the SM Higgs coupling. This choice can avoid the 
constraints from the signal data of the 125 GeV Higgs. 

{\bf (3) Searches for extra Higgses at the LHC and flavor physics constraints.}
The Yukawa couplings of the other Higgses ($H$, $H^\pm$, $A$, $\bm{a}$) are proportional 
to $\kappa_u$, $\kappa_d$ and $\kappa_\ell$ when $\sin(\beta-\alpha) \approx 1$. Therefore, we assume $\kappa_u$, $\kappa_d$ and $\kappa_\ell$ to be small enough to  
suppress the production cross sections of these Higgses at the LHC, and consistent with the exclusion limits of searches for additional Higgs bosons.
Additionally, very small values of $\kappa_u$, $\kappa_d$, and $\kappa_\ell$ satisfy the bounds from flavor physics observables.
In other words, we assume that the Yukawa couplings can be tuned so that the searches for extra Higgses at the LHC and flavor physics constraints are truly absent.

{\bf (4) The oblique electroweak parameters.}
The oblique parameters $(S, T, U)$ quantify the effects of electroweak symmetry breaking. Additional Higgs particles introduce new Feynman diagrams through interactions with $W$ and $Z$ bosons, impacting the $S$, $T$ and $U$ parameters. The expressions for $S,\;T$ and $U$ in this model are \cite{stu1,stu2}
\begin{equation}
	\begin{aligned}
			S = &\frac{1}{{\pi m_Z^2}}\left[ {c_\theta ^2{F_S}(m_Z^2,m_H^2,m_A^2) + s_\theta ^2{F_S}(m_Z^2,m_H^2,m^2_{\bm{a}}) - {F_S}(m_Z^2,m_{{H^ \pm }}^2,m_{{H^ \pm }}^2)} \right],\\
			T =& \frac{1}{{16\pi m_W^2s_W^2}}\left[ { - c_\theta ^2{F_T}(m_H^2,m_A^2) - s_\theta ^2{F_T}(m_H^2,m^2_{\bm{a}}) + {F_T}(m_{{H^ \pm }}^2,m_H^2)} \right.\\
			{\rm{      }} 
			&+ \left. {c_\theta ^2{F_T}(m_{{H^ \pm }}^2,m_A^2) + s_\theta ^2{F_T}(m_{{H^ \pm }}^2,m^2_{\bm{a}})} \right],\\
			U = &\frac{1}{{\pi m_W^2}}\left[ {{F_S}(m_W^2,m_{{H^ \pm }}^2,m_H^2) - 2{F_S}(m_W^2,m_{{H^ \pm }}^2,m_{{H^ \pm }}^2)} \right.\\
			&\left. {{\rm{      }} + c_\theta ^2{F_S}(m_W^2,m_{{H^ \pm }}^2,m_A^2) + s_\theta ^2{F_S}(m_W^2,m_{{H^ \pm }}^2,m^2_{\bm{a}})} \right]\\
			& - \frac{1}{{\pi m_Z^2}}\left[ {c_\theta ^2{F_S}(m_Z^2,m_H^2,m_A^2) + s_\theta ^2{F_S}(m_Z^2,m_H^2,m^2_{\bm{a}})} \right.\left. { - {F_S}(m_Z^2,m_{{H^ \pm }}^2,m_{{H^ \pm }}^2)} \right],
	\end{aligned}
\end{equation}
where
\begin{equation}
		\begin{aligned}
	&{F_T}(a,b) = \frac{1}{2}(a + b) - \frac{{ab}}{{a - b}}\log (\frac{a}{b}),\\
	&{F_S}(a,b,c) = {B_{22}}(a,b,c) - {B_{22}}(0,b,c),\\
	&{B_{22}}(a,b,c) = \frac{1}{4}\left[ {b + c - \frac{1}{3}a} \right] - \frac{1}{2}\int_0^1 dx X\log (X - i),\\
	&X = bx + c(1 - x) - ax(1 - x).
		\end{aligned}
\end{equation}
Based on the recent fitting results of Ref.\cite{pdg2020}, we use the following values of $S$ , $T$ and $U$ 
\begin{equation}
	S =  - 0.01 \pm 0.10,\;T =  - 0.03 \pm 0.12,\;U = 0.02 \pm 0.11,
\end{equation}
with the correlation coefficients 
\beq
\rho_{ST} = 0.92,~~  \rho_{SU} = -0.80,~~  \rho_{TU} = -0.93.
\eeq
\section{ELECTROWEAK PHASE TRANSITION}\label{sec4}
We utilized the publicly available code \textbf{PhaseTracer} \cite{phasetracer,phasetracer2} and \textbf{EasyScan\_HEP} \cite{easyscan} for numerical analyses of PTs. \textbf{PhaseTracer} traces the minima of the effective potential as temperature varies and calculates the critical temperatures $T_{c}$ where these minima become degenerate. This approach helps us investigate the complex dynamics of PTs involving multiple scalar fields. We use $\gamma$ as a metric to quantify the strength of a PT \cite{Ahriche:2007jp,Ahriche:2014jna}, defined as: 
\begin{equation}
	\begin{aligned}
	&\gamma  \equiv \frac{{v\left( {T_{c}} \right)}}{{T_{c}}},		
	\end{aligned}
\end{equation}
where $v = \sqrt {h_1^2 + h_2^2+h_3^2 + h_4^2}$ represents the VEV at true vacuum.
To ensure that the current universe is in the observed electroweak symmetry breaking vacuum, we necessitate that the vacuum characterized by $v = \sqrt {h_1^2 + h_2^2}  = 246\;{\rm{GeV}}$ and $h_3=h_4=0$ is the deepest vacuum at zero temperature. The metastable electroweak symmetry breaking vacuum with a lifetime exceeding the age of the universe is not within the scope of this discussion. 

\begin{figure}[tb]
	\centering
		\includegraphics[page=1, width=\textwidth]{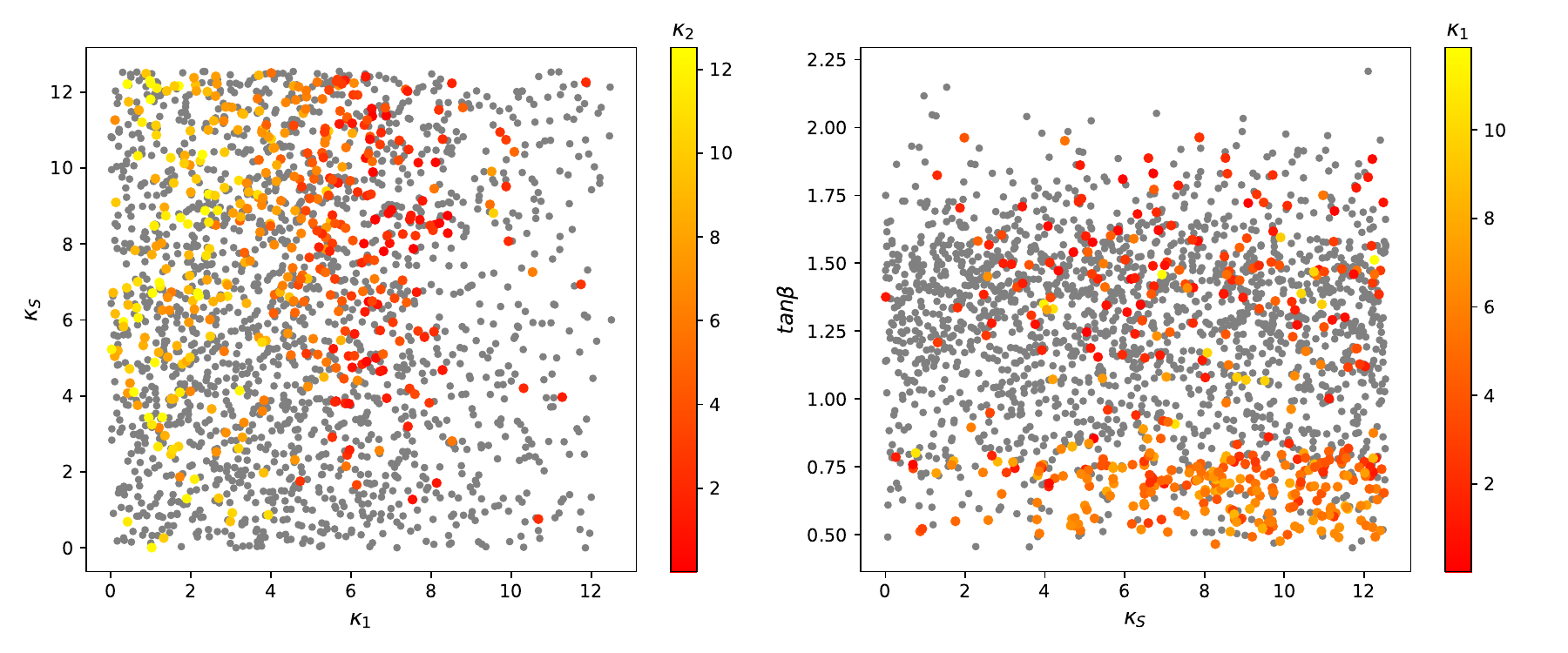}
        \caption{The viable parameter regions for FOPT. Both the gray and colored scatter points satisfy the theoretical and experimental constraints, but only the colored scatter points correspond to FOPT.}	
	\label{scan}
\end{figure}

We conducted a random uniform scan over the parameter region to identify the range of FOPT most likely to occur,
\begin{equation}
    \begin{aligned}
    &300~{\rm{GeV}} <m_{H,H{\pm}} < 600~{\rm{GeV}},
    ~~ 140~{\rm{GeV}} <m_A < 300~{\rm{GeV}}, \\
    &400~{\rm{GeV}} <m_{\bm{a}} < 600~{\rm{GeV}},
    ~~0 < \kappa_{1,2,S} < 4\pi, 
    ~~0.2 < \rm{tan}\beta < 5.0,\\
    & -5000~{\rm{GeV^2}} < m^2_{12} <5000~{\rm{GeV^2}},
    ~~-0.6 < \rm{sin}\theta < 0.6.
    \end{aligned}
\end{equation}
In FIG.\ref{scan}, the colored scatter points correspond to the surviving samples that have FOPT and satisfy the theoretical and experimental constraints discussed in Section \ref{sec3}. The gray scatter points are excluded because there is no FOPT but fulfilled the theoretical and experimental constraints.
From the left panel of FIG.\ref{scan}, $\kappa_{1}$, $\kappa_{2}$, or $\kappa_{S}$ are allowed to take very small values, but they cannot all be very small simultaneously. The right panel shows that most of the surviving samples have tan$\beta$ values ranging from 0.5 to 2.0. Furthermore, surviving samples with tan$\beta > $1.0 do not exhibit spontaneous CP violation, and the PT strength is less than 1.
Given our focus on the phenomenology of strong first-order electroweak PT, we have selected benchmark points (BPs) with tan$\beta < $ 1.0. These benchmark points, characterized by larger values of $\kappa_{1,2,S}$, are chosen based on the premise that such parameters yield a stronger GW spectrum, making them testable by space-based GW detectors.

We select five BPs in the surviving parameter space to detail the physical processes associated with the PTs. The PT histories are illustrated in FIG. \ref{BP1} to FIG. \ref{BP5}, which shows the VEVs of the scalar fields as functions of temperature. Arrows in the figures indicate that at the critical temperatures, the two phases connected by the arrows are degenerate, implying the possibility of a first-order PT occurring along the direction of the arrow. Dots represent the vacuum fields that remain unchanged during the transition.
The parameters for these BPs are provided in TABLE \ref{BPPP}, and $T_{c}$ and PT strengths at each step are shown in TABLE \ref{strength}.

The PT types mentioned in Ref.\cite{Liu:2023sey}, using high temperature approximation where CP symmetry is broken at high temperatures and restored at low temperatures,  align with PB3 and BP4 found in our work. Additionally, we have identified two unique PT types, as specifically illustrated in FIG. \ref{BP2} and FIG. \ref{BP5}.
In the subsequent discussion, we categorize these PT processes into two distinct classes, based on whether the CP-symmetry is broken at a finite temperature.

\setlength{\arrayrulewidth}{1.0pt}
\begin{table}
\resizebox{\textwidth}{!}{ 
	\begin{tabular}{|| c || c || c || c || c || c || c || c || c || c || c ||}
		\hline
		& $m_H$(GeV) &$m_A$(GeV)& $m_{\bm{a}}$(GeV) & $m_{0}^2$(GeV$)^2$ & $\kappa_1$ & $\kappa_2$ & $\kappa_S$ &$\mu$ & $\tan\beta$  \\
		\hline
		\hline  
		$\textbf{BP1}$ & ~~333.76~~ & ~~293.99~~ & ~~416.07~~ & ~~-540.30~~& ~~6.52~~ & ~~3.34~~ & ~~6.49~~ &~~-118.46~~& ~~0.75~~\\  
		\hline
		$\textbf{BP2}$ & ~~309.13~~ & ~~148.09~~ & ~~474.45~~ & ~~-999.19~~  & ~~3.17~~ & ~~7.83~~ & ~~7.76~~ & ~~-366.25~~& ~~0.88~~\\  
		\hline 
		$\textbf{BP3}$ & ~~333.69~~ & ~~205.13~~ & ~~406.42~~ & ~~-2734.90~~ & ~~4.91~~ & ~~3.48~~ & ~~10.39~~ &~~-219.74~~& ~~0.68~~\\ 
		\hline  		
        $\textbf{BP4}$ & ~~353.55~~ & ~~152.49~~ & ~~414.22~~ & ~~-1892.17~~ & ~~2.56~~ & ~~9.08~~ & ~~11.24~~ &~~-272.67~~& ~~0.56~~\\  
		\hline  
        $\textbf{BP5}$ & ~~381.65~~ & ~~179.28~~ & ~~422.95~~ & ~~-12309.84~~ & ~~3.06~~ & ~~7.45~~ & ~~10.17~~ &~~-263.37~~& ~~0.81~~\\  
		\hline 
	\end{tabular}
 }
\caption{The input parameters for the BPs. $m_H=m_{H^{\pm}}$ are chosen to satisfy the constraints from the oblique parameters.}
\label{BPPP}
\end{table}

\begin{table}
	\begin{tabular}{|| c || c || c || c || c || c || c ||}
		\hline
		& $T_{c1}$ &$\gamma_1$& $T_{c2}$ &$\gamma_2$ & $T_{c3}$ & $\gamma_3$   \\
		\hline
		\hline 
		$\textbf{BP1}$ &~~106.57~~    & ~~1.58~~ & ~~-~~        & ~~-~~ & ~~-~~      & ~~-~~     \\  
		\hline
		$\textbf{BP2}$ &~~168.36~~    & ~~0.38~~ & ~~98.31~~        & ~~1.58~~ & ~~-~~      & ~~-~~     \\  
		\hline
		$\textbf{BP3}$ &~~136.67~~    & ~~0.47~~ & ~~99.29~~   & ~~2.01~~ & ~~-~~      & ~~-~~     \\  
		\hline
		$\textbf{BP4}$ &~~146.11~~    & ~~0.43~~ & ~~77.79~~   & ~~2.91~~ & ~~-~~      & ~~-~~     \\ 
		\hline		
	$\textbf{BP5}$ &~~114.12~~    & ~~0.93~~ & ~~104.16~~    & ~~1.83~~ & ~~103.35~~      & ~~1.86~~     \\  	  
		\hline
	\end{tabular}
\caption{Critical temperatures and strengths of the PTs for BPs. }
\label{strength}
\end{table}

\subsection{{CP symmetry conserved at all temperatures.}} 	

\begin{figure}[tb]
	\centering
	\begin{minipage}{0.45\textwidth}
		\includegraphics[page=1, width=\textwidth]{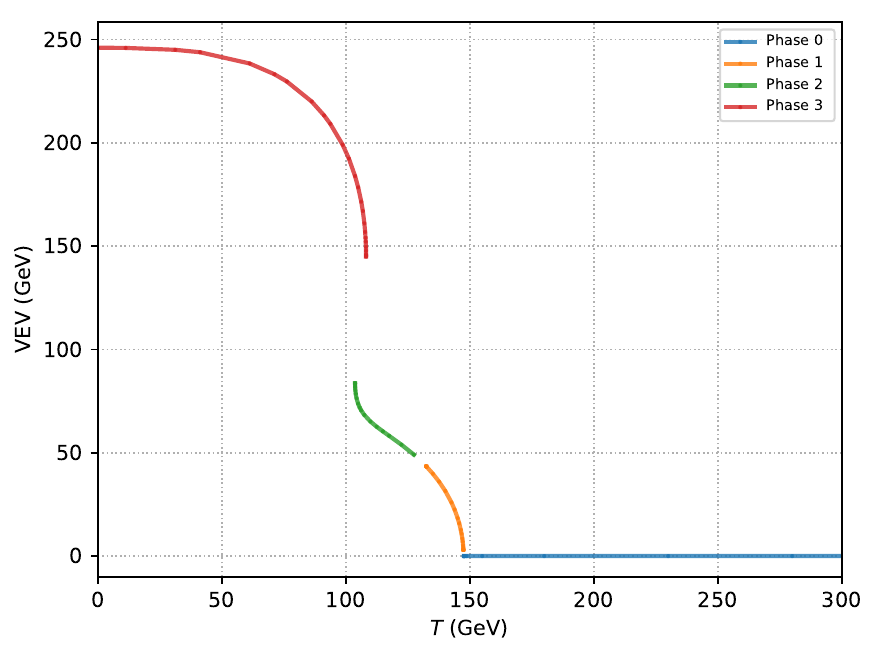}
	\end{minipage}
	\begin{minipage}{0.45\textwidth}
		\includegraphics[page=2, width=\textwidth]{h1h2-nocpv-1.pdf}
	\end{minipage}
         \caption{Phase histories for BP1. The black arrows depict the critical temperatures, hinting at the possibility of a FOPT occurring in the direction of the arrows. The dots, on the other hand, signify that the corresponding vacuum field remains unchanged during the transition.}	
	\label{BP1}
\end{figure}

\begin{figure}[tb]
	\centering
	\begin{minipage}{0.95\textwidth}
		\includegraphics[page=1, width=\textwidth]{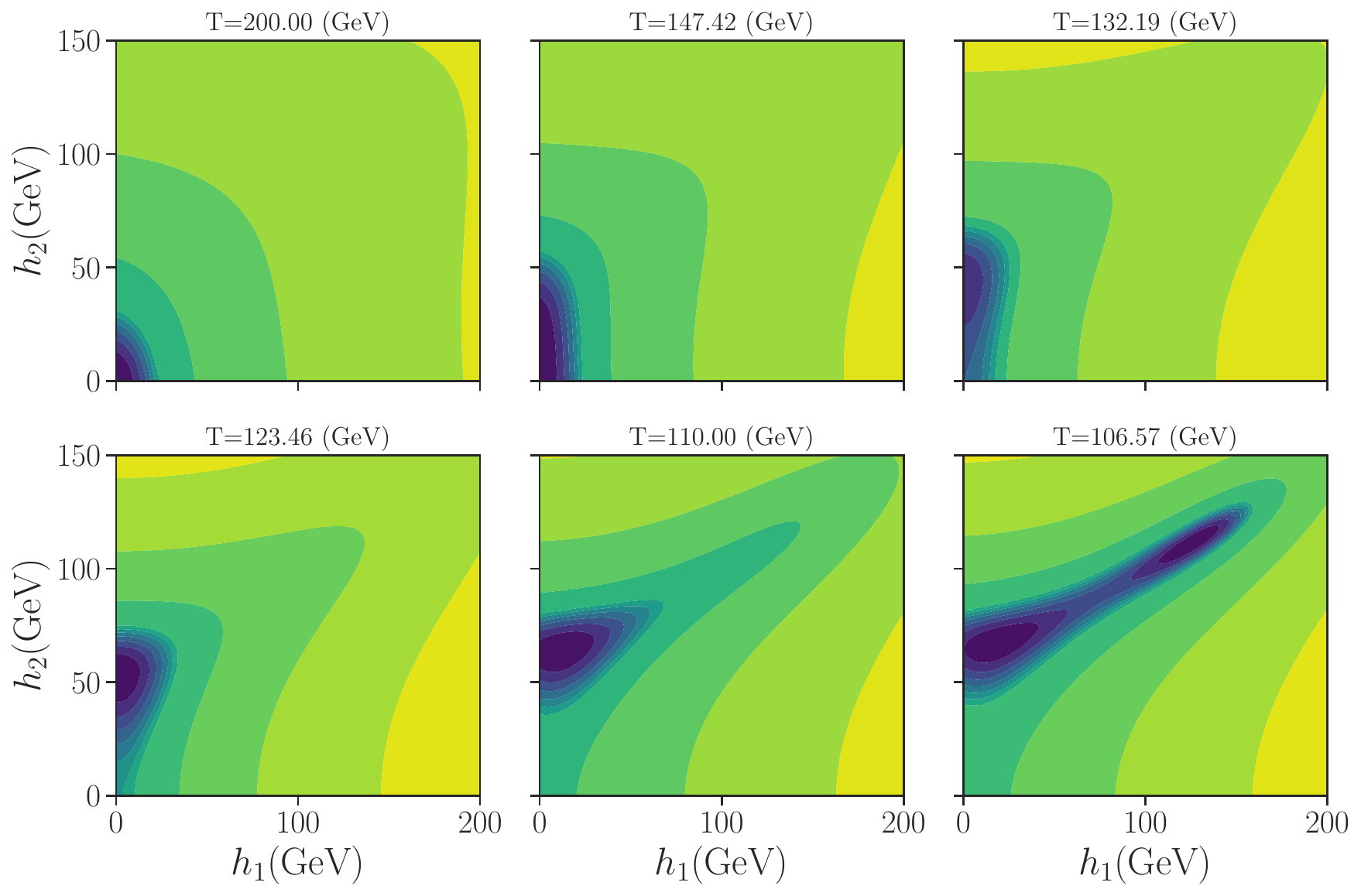}
	\end{minipage}
    \caption{Effective potentials at different temperatures for BP1, with $h_3=h_4=0$ GeV. The color represents the value of the effective potential, with darker shades corresponding to lower potential values.}	
	\label{BP1_p}
\end{figure}

For the scenarios described by BP1, only the field configurations $h_1$ and $h_2$ acquire non-zero VEVs and break the electroweak symmetry, while the VEVs of other fields remain zero. So under these scenarios, the CP symmetry is conserved in the whole thermal history. 
In the BP1 scenario, electroweak symmetry is firstly broken through a cross-over transition, characterized by the continuous evolution of the effective potential's minima, as shown in FIG. \ref{BP1} and FIG. \ref{BP1_p}, illustrating the shift from Phase 0 to Phase 2. This kind of smooth-crossover PT can be accommodated with the vary of scalar minima in the FIG. \ref{BP1_p} from $T=200.00$ GeV to $T=110.00$ GeV. The darkest range indicates the location of the global minima.
The system evolves in Phase 2 until the temperature drops to 106.57 GeV, and then the minima of the effective potential has two different values in Phase 2 and Phase 3 respectively. Consistent with the last panel of FIG. \ref{BP1_p}, the barrier between the respective minima becomes apparent. A strong first-order electroweak PT takes place and the PT strength is greater than one.


\subsection{{ CP symmetry broken at high temperatures and recovered at the present temperatures.}}
Unlike the scenario discussed above, we now turn to the special cases described by BP2 to BP5. In these cases, the CP-odd Higgs fields $h_3$ and $h_4$ acquire nonzero VEVs at high temperatures but return to zero as the temperature decreases. Thus, the CP symmetry is spontaneously broken at high temperature and then restored at the present temperatures.
\begin{itemize}
	\item {Electroweak and CP symmetries broken in sequence.}
\end{itemize}

\begin{figure}[tb]
	\centering
	\begin{minipage}{0.45\textwidth}
		\includegraphics[page=1, width=\textwidth]{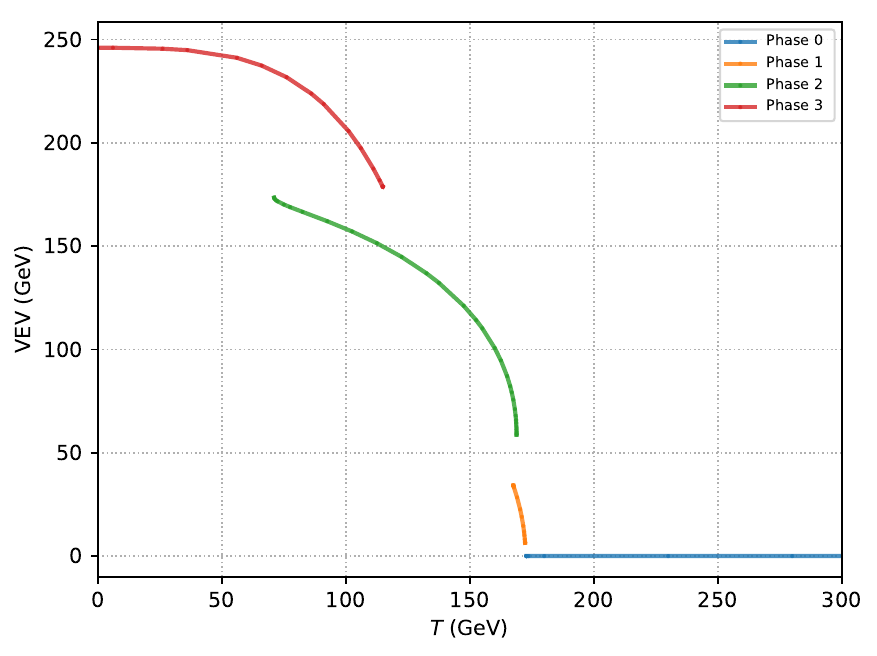}
	\end{minipage}
	\begin{minipage}{0.45\textwidth}
		\includegraphics[page=2, width=\textwidth]{h1h2.pdf}
	\end{minipage}
	\caption{Same as Fig.~\ref{BP1}, but for BP2.}
	\label{BP2}
\end{figure}

The subsequent section will provide a comprehensive description of the BP2 process. As shown in the right panel of FIG. \ref{BP2}, in Phase 1 $h_1$ and $h_2$ acquire non-zero VEVs through a cross-over PT and break the electroweak symmetry spontaneously, while the CP symmetry remains conserved since the configurations of CP-odd Higgs fields do not acquire non-zero VEVs.
However, when the temperature decreases to 168.36 GeV, the CP symmetry is broken as $h_3$ and $h_4$ acquire non-zero VEVs. This marks a first-order electroweak PT from Phase 1 to Phase 2. 
Subsequently, as the temperature drops further to 98.31 GeV, the system tunnels from Phase 2 to Phase 3 via a strong first-order electroweak PT. The VEVs of the four fields $(\left\langle {{h_1}} \right\rangle,\left\langle {{h_2}} \right\rangle,\left\langle {{h_3}} \right\rangle,\left\langle {{h_4}} \right\rangle)$ shift from $(104.32,46.26,92.04,62.08)$ GeV to $(151.88,144.62, 0.0, 0.0)$ GeV.
In this final phase, the CP symmetry is recovered. The system continues to evolve along the last phase, eventually settling into the observed vacuum state.

\begin{itemize}
	\item {Electroweak and CP symmetry broken Together}
\end{itemize}

\begin{figure}[tb]
	\centering
	\begin{minipage}{0.45\textwidth}
		\includegraphics[page=1, width=\textwidth]{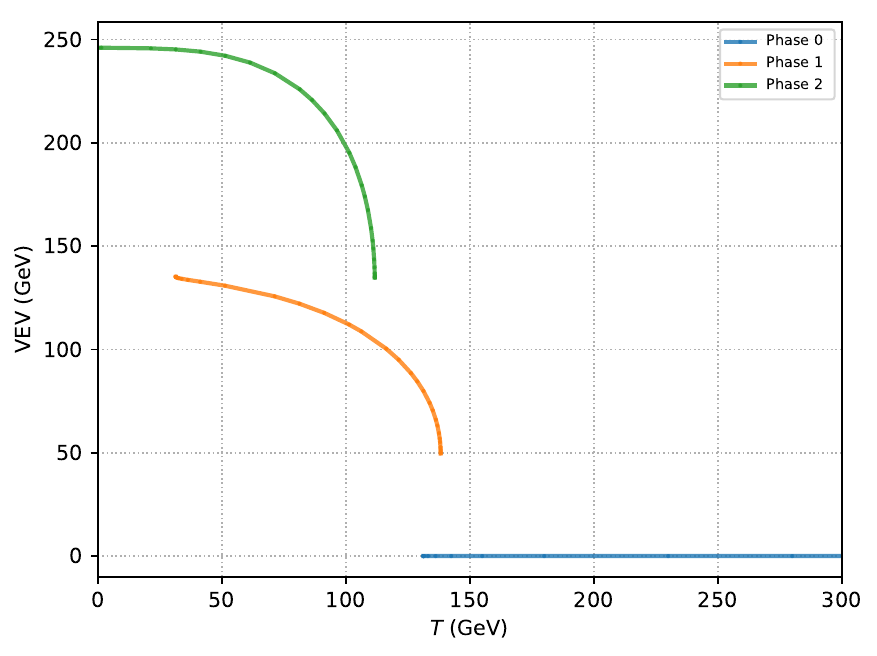}
	\end{minipage}
	\begin{minipage}{0.45\textwidth}
		\includegraphics[page=2, width=\textwidth]{h1h2h3h4.pdf}
	\end{minipage}
	\caption{Same as Fig.~\ref{BP1}, but for BP3.}
	\label{BP3}
\end{figure}
The PT process of BP3 is similar to the second and third steps of the BP2 scenario. A degenerate state appears between Phase 0 and Phase 1 at the critical temperature $T_{c1}$. After this, $h_1$, $h_2$, $h_3$ and $h_4$ acquire non-zero VEVs, indicating a new phase where both electroweak and CP symmetries are broken. 
In Phase 1, the VEVs of the four fields gradually increase as the temperature decreases. The second PT occurs at $T_{c2}$ = 99.29 GeV, at which the system undergoes a strong first-order electroweak PT, tunneling into the final phase and restoring the CP symmetry.

\begin{itemize}
	\item {CP symmetry broken before electroweak symmetry.}
\end{itemize}

\begin{figure}[tb]
	\centering
	\begin{minipage}{0.45\textwidth}
		\includegraphics[page=1, width=\textwidth]{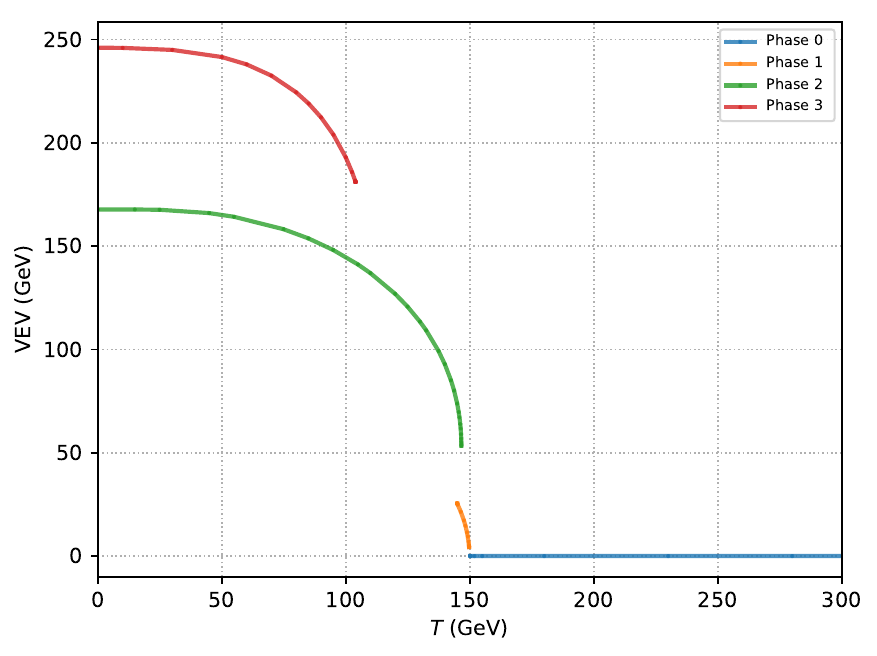}
	\end{minipage}
	\begin{minipage}{0.45\textwidth}
		\includegraphics[page=2, width=\textwidth]{h4-1.pdf}
	\end{minipage}
\caption{ Same as Fig.~\ref{BP1}, but for BP4. 
} 
\label{BP4}
\end{figure}
\begin{figure}[tb]
	\centering
	\begin{minipage}{0.45\textwidth}
		\includegraphics[page=1, width=\textwidth]{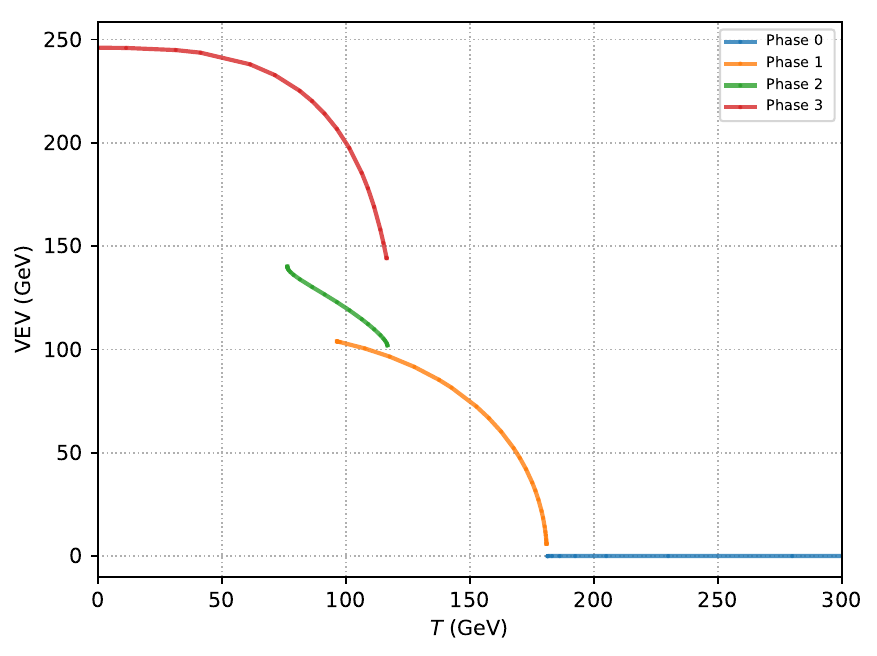}
	\end{minipage}
	\begin{minipage}{0.45\textwidth}
		\includegraphics[page=2, width=\textwidth]{h4-2.pdf}
	\end{minipage}
	\caption{Same as Fig.~\ref{BP1}, but for BP5}
	\label{BP5}
\end{figure}
For the BP4 the Universe undergoes four different phases, and a cross-over PT takes place from Phase 0 to Phase 1.
In Phase 1, the CP symmetry is broken spontaneously when the $h_4$ field acquires a non-zero VEV while the VEVs of $h_1$ and $h_2$ remain zero, respecting electroweak symmetry conservation. In Phase 2, the remaining fields, especially $h_1$ and $h_2$, acquire non-zero VEVs and break the electroweak symmetry spontaneously.  
Phase 2 represents an unstable false vacuum and the system will tunnel into the final phase by a strong first-order electroweak PTs. 
Phase 3 is the final phase, where the VEVs of $h_3$ and $h_4$ return to zero, restoring the CP symmetry. Ultimately, the observed vacuum is obtained as the temperature approaches zero.

BP5 exhibits two potential transition pathways: (i) sequential transitions, proceeding first from Phase 1 to Phase 2, and subsequently from Phase 2 to Phase 3, and (ii) direct transitions from Phase 1 to Phase 3, bypassing Phase 2. This is quite distinct from previous benchmark points and may yield a rich phenomenology. The calculation of their transition probabilities requires special consideration and is left for future research.

\section{GRAVITATIONAL WAVE}\label{sec5}
During the first-order PT, the GW can be generated, which has three primary sources: bubble collisions, plasma sound waves, and magnetohydrodynamic turbulence. 
Bubble collisions becomes significant when the bubble wall velocity $\tilde{v}_W$ approaches the speed of light. Here we focus on plasma sound waves and magnetohydrodynamic turbulence as the dominant contributors to GW production in our subsequent calculations. 

The contribution originating from the sound waves can be expressed by \cite{gw-sw}
\begin{eqnarray}
\Omega_{\textrm{sw}}h^{2} & \ = \ &
2.65\times10^{-6}\left( \frac{H(T_{*})}{\beta}\right)\left(\frac{\kappa_{\rm sw} \alpha}{1+\alpha} \right)^{2}
\left( \frac{100}{g_{\ast}(T_{*})}\right)^{1/3} \tilde{v}_W\nonumber \\
&&\times  \left(\frac{f}{f_{\rm sw}} \right)^{3} \left( \frac{7}{4+3(f/f_{\textrm{sw}})^{2}} \right) ^{7/2} \Upsilon(\tau_{\rm sw})\ ,
\label{eq:soundwaves}
\end{eqnarray}
$H(T_{*})$ is the Hubble parameter at reference temperature $T_{*}$. $g_*(T_*)$ is the effective number of relativistic degrees of freedom at $T_*$.
The $\kappa_{\rm sw}$ is the fraction of latent heat transformed into the kinetic energy of the fluid \cite{1004.4187}. The suppression factor $\Upsilon(\tau_{\rm sw})$ is a function of the lifetime of the source \cite{2007.08537}.
\begin{equation}
\Upsilon(\tau_{\rm sw})=1-\frac{1}{\sqrt{1+2\tau_{\rm sw}H_n}}\ ,
\end{equation}
arises due to the finite lifetime $\tau_{sw}$ of the sound waves \cite{2003.07360,2003.08892},
\begin{equation}
\tau_{\rm sw}=\frac{\tilde{v}_W(8\pi)^{1/3}}{\beta \bar{U}_f}, ~~ \bar{U}^2_f=\frac{3}{4}\frac{\kappa_{\rm sw}\alpha}{1+\alpha}\ .
\end{equation}
$\alpha$ is the transition strength parameter, defined as the ratio of the vacuum energy density released during the PT to the total radiation energy density at the reference temperature.
 \begin{equation}
      \alpha=\frac{1}{\pi^2g_*(T)T^4/30}\left( {\Delta {V_{{\rm{eff}}}} - \frac{T}{4}\frac{{d\Delta {V_{{\rm{eff}}}}}}{{dT}}} \right)|_{T=T_*},
  \end{equation}  
$\beta$ represents the inverse of the time duration of the PT,
\begin{equation}
\beta=TH(T)\frac{d (S_3/T)}{d T}|_{T=T_*}\; .
\end{equation}
$S_3$ is the three-dimensional Euclidean action. $f_{\text{sw}}$ is the present peak frequency of the spectrum,
\begin{equation}
f_{\textrm{sw}} \ = \ 
1.9\times10^{-5}\frac{1}{\tilde{v}_W}\left(\frac{\beta}{H(T_{*})} \right) \left( \frac{T_{*}}{100\textrm{GeV}} \right) \left( \frac{g_{\ast}(T_*)}{100}\right)^{1/6} \textrm{Hz} \,.
\label{fsw}
\end{equation}

The contribution from the turbulence can be expressed by \cite{gw-turb_1,gw-turb_2}
\begin{eqnarray}
\Omega_{\textrm{turb}}h^{2} & \ = \ &
3.35\times10^{-4}\left( \frac{H(T_{*})}{\beta}\right)\left(\frac{\kappa_{\rm turb} \alpha}{1+\alpha} \right)^{2}
\left( \frac{100}{g_{\ast}(T_*)}\right)^{1/3} \tilde{v}_W\nonumber \\
&&\times \frac{\left({f}/{f_{\rm turb}}\right)^{3}} {\left(1+f/f_{\rm turb}\right)^{11/3}\left(1+8\pi f/H_{0}\right)},
\label{eq:turb}
\end{eqnarray}
we take $\kappa_{\rm turb} \approx 0.1~ \kappa_{\rm sw}$ and the $H_{0}$ is
\begin{equation}
H_{0} \ = \ 
1.65\times10^{-5}\left( \frac{T_{*}}{100\textrm{GeV}}\right ) \left( \frac{g_{\ast}(T_*)}{100}\right)^{1/6}\,.
\label{fturb}
\end{equation}
The present peak frequency
\begin{equation}
f_{\textrm{turb}} \ = \ 
2.7\times10^{-5}\frac{1}{\tilde{v}_W}\left(\frac{\beta}{H(T_{*})} \right) \left( \frac{T_{*}}{100\textrm{GeV}} \right) \left( \frac{g_{\ast}(T_*)}{100}\right)^{1/6} \textrm{Hz} \,.
\end{equation}

In Section \ref{sec4} we set the reference temperature to the critical temperature $T_{c}$ to analyze multi-step PTs. Bubbles form in the unstable false vacuum when the universe's temperature drops below $T_{c}$. 
The nucleation temperature $T_{n}$ is commonly regarded as the starting point of the phase transition, defined as the moment when the probability of nucleating a supercritical bubble within one Hubble volume reaches approximately one \cite{Hindmarsh:2020hop}.
\begin{equation}
\int_{{T_{c}}}^{{T_{n}}} {\frac{{dT}}{T}} {\left( {\frac{{90}}{{{\pi ^2}{g_*}\left( T \right)}}} \right)^2}{\left( {\frac{{{m_{\rm pl}}}}{T}} \right)^4}{e^{ - {S_3}\left( T \right)/T}} = \mathcal{O}(1).
\label{tn-cond}
\end{equation}
We use $T_{n}$ as the reference temperature $T_{*}$ to examine the production of the stochastic GW background.
Bubble nucleation is a stochastic process and the nucleation rate per unit volume and time is given by $\Gamma=Ae^{-S_3/T}$, where $A$ is typically approximated as the fourth power of the temperature at high temperatures. 

In our discussion of first-order fast PT at the electroweak scale, the integrals can be roughly approximated as being dominated by their value at $T_{n}$. $g_*(T_{n})\approx106.75$ and $m_{\rm pl}$ is the reduced Planck mass in the natural system of units, i.e. $m_{\rm pl}=2.4\times10^{18}$ GeV. Thus, Eq.(\ref{tn-cond}) is approximately satisfied for $S_3(T_{n})$/$T_{n}\approx140$. 
 In spherical coordinates, $S_3$ is given by~\cite{Linde:1980tt},
\begin{eqnarray}
    {S_3} = 4\pi \int_o^\infty  {dr{r^2}\left[ {\sum\limits_{i = 1}^4 {\frac{1}{2}{{\left( {\frac{{d{h_i}}}{{dr}}} \right)}^2} + } {\rm{ }}{V_{eff}}} \right]}.
\end{eqnarray}
The field configurations can be solved via the bounce equations,
\begin{equation}
\label{eq: bubble_equation}
\frac{ \mathrm{d}^2 h_i }{ \mathrm{d} r^2 } + \frac{ 2 }{ r } \frac{ \mathrm{d} h_i }{ \mathrm{d} r } = \frac{ \partial V_{\rm eff} }{ \partial h_i }, \quad ( i=1,2,3,4). 
\end{equation}
We use the publicly code \textbf{CosmoTransitions} \cite{cosmotransition} to compute $S_3$. TABLE \ref{Tn} lists the $T_{n}$, the corresponding $\alpha$ and $\beta/H$ for various BP scenarios. When the difference between $T_c$ and $T_n$ is relatively small, the phase transition is very fast. In such cases, the calculated $\beta/H$ are too large and therefore not displayed in the table.

The fluid motion can be classified into three modes based on the relationship between the bubble wall velocity and the speed of sound: weak detonation, hybrid, and weak deflagration. Different modes of fluid motion have a significant impact on the results of GW signals. To illustrate these effects, three distinct cases are considered: ${v_W} = 0.4$, ${v_W} = 0.6$, and ${v_W} = 0.9$. The bubble wall velocities are chosen to correspond to the three fluid motion modes, as determined by the bag model \cite{bagmodel}.

\begin{table}
	\centering
	\begin{tabular}{ c p{2.2cm}  p{2.2cm} p{2.2cm} p{2.2cm} p{2.2cm} p{2.2cm} p{2.2cm}  }
		\hline
		\hline
		           &~BP1~~~      &~BP2         &~BP3       &~BP4        &~BP5\\
		\hline 
		~~~~$T_{n1}~~$&105.58    &168.20       &136.08      &145.99    &113.28 \\
		$\alpha$      &0.0079   &0.0005       &0.0008      &0.0006         &0.0003  \\
		$\beta/H$     &26313.36   &595647.81      &59188.97   &~~~~-          &~~~~- \\
		\hline
		$T_{n2}$     &~~~~      &89.02   &85.77    &52.18        & 97.40\\
		$\alpha$     &~~~~      &0.0109   &0.0210       &0.0726          &0.0126 \\
		$\beta/H$    &~~~~      &2588.97  &1263.01     &254.30        &2149433.20\\
		\hline
		$T_{n3}$     &~~~~      &~~~~      &~~~~       &~~~~          & 94.22  \\
		$\alpha~$    &~~~~      &~~~~      &~~~~      &~~~~          & 0.0133\\
		$\beta/H$    &~~~~      &~~~~      &~~~~       &~~~~          & 2651.97  \\
		\hline
		\hline
	\end{tabular}
	\caption{Nucleation temperatures, $\alpha$ and $\beta/H$ corresponding to different steps in the phase transition of BP1 to BP5. Here, the corresponding $\beta/H$ indicated by ` - ' is too large to be numerically obtained in practice. }
\label{Tn}
\end{table}

\begin{figure}[htbp!]
    \centering
    \includegraphics[width=0.49\textwidth]{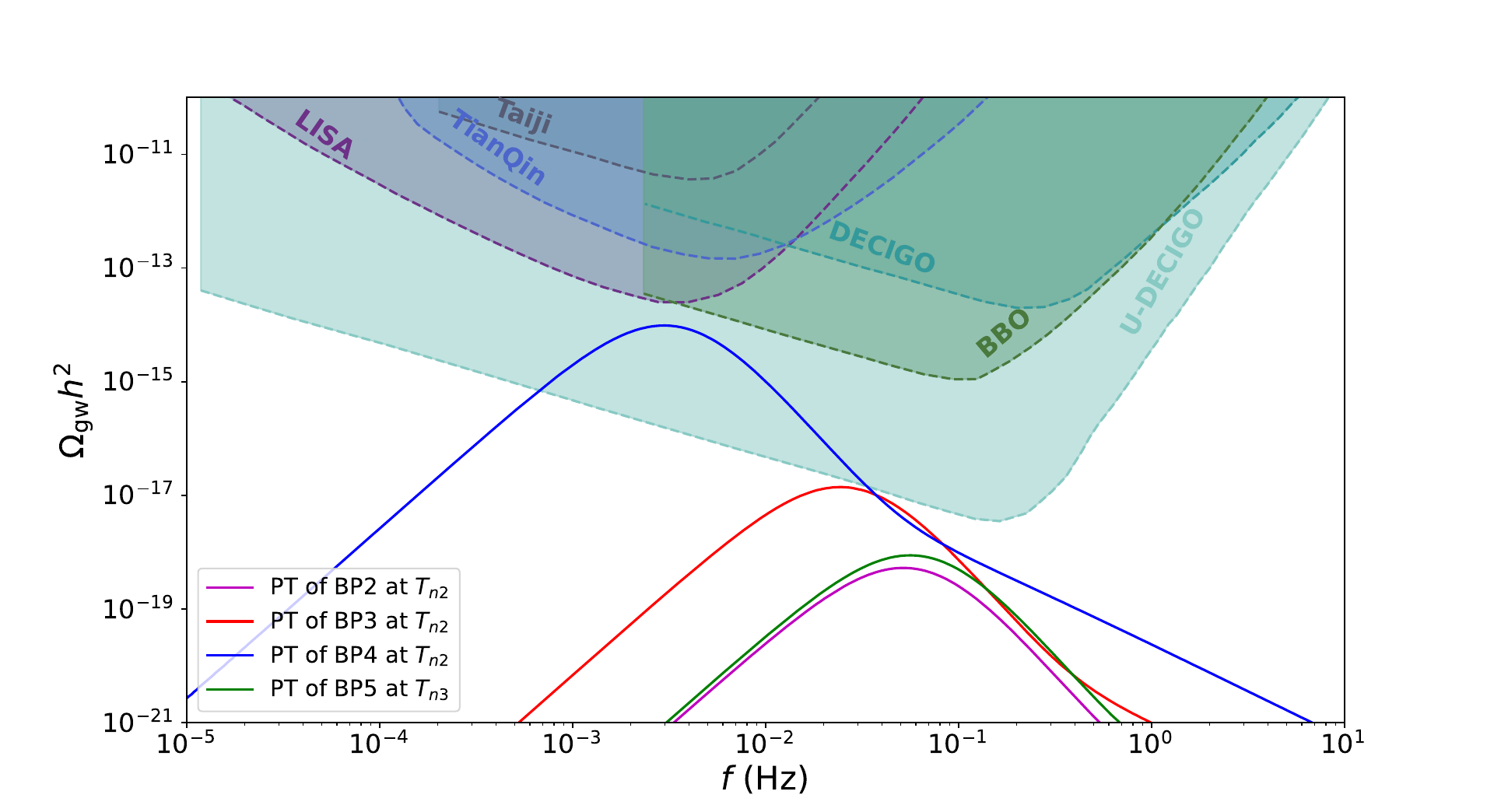}
    \includegraphics[width=0.49\textwidth]{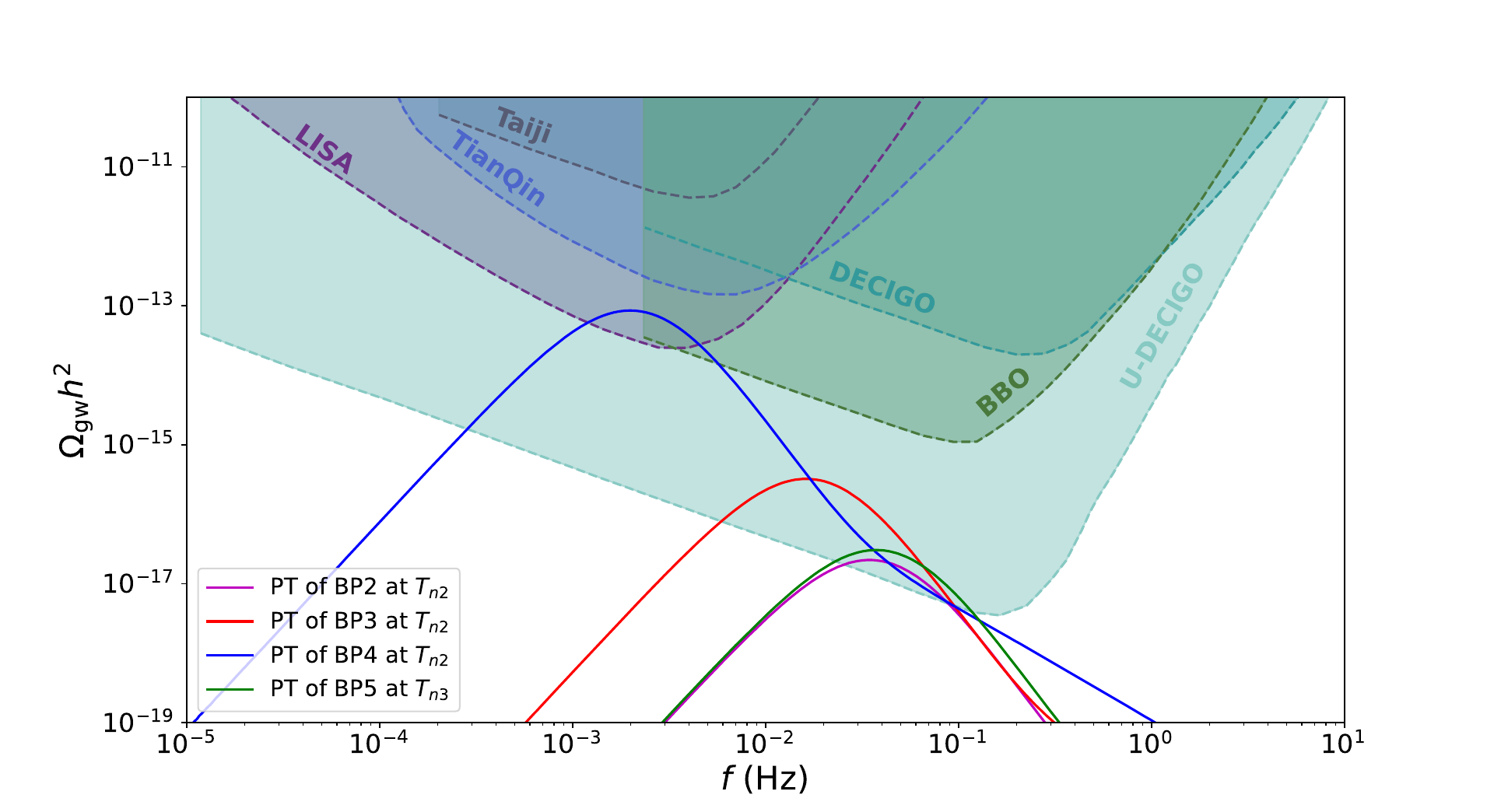}
    \includegraphics[width=0.5\textwidth]{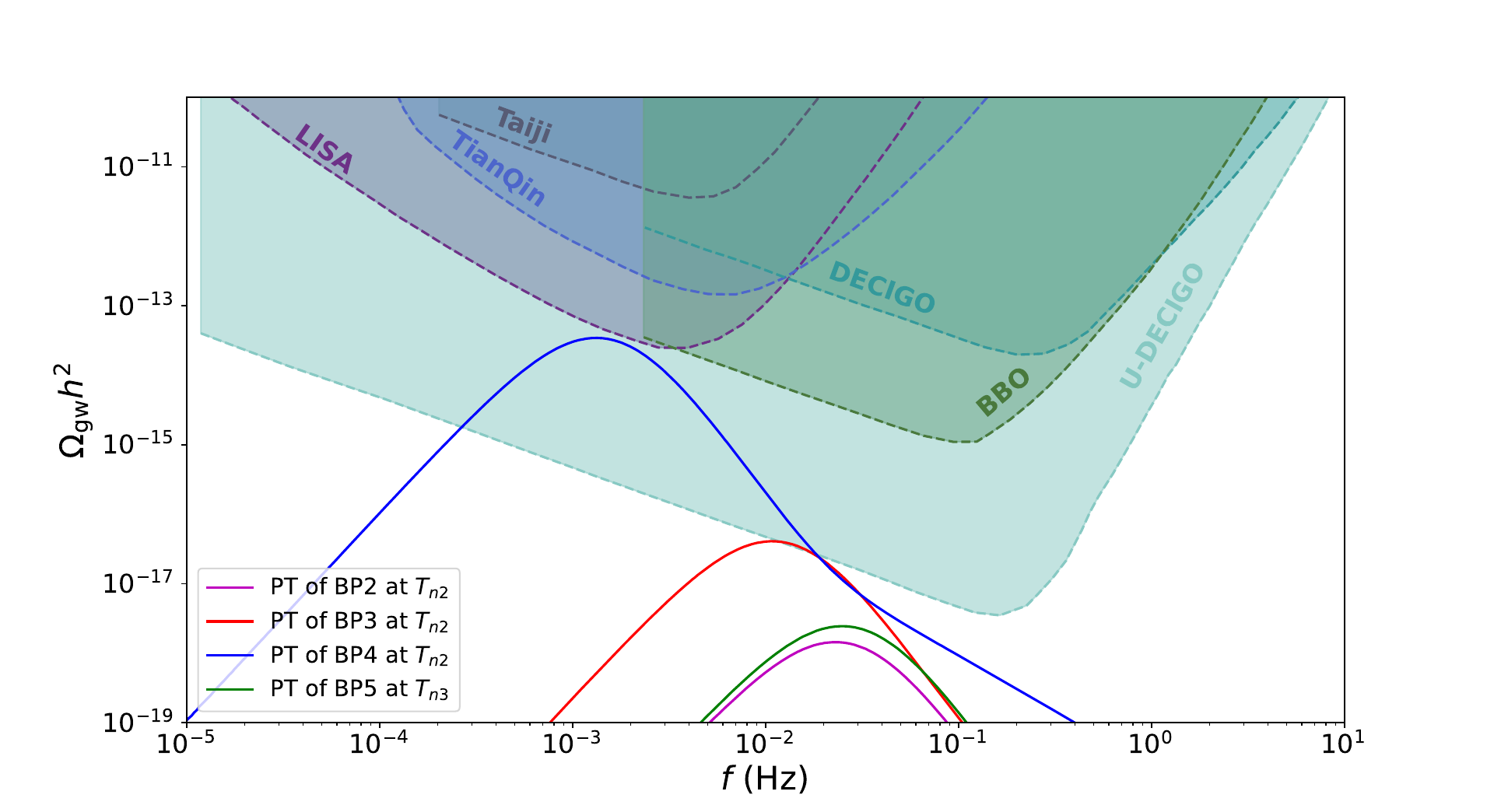}
    \caption{
    Gravitational wave spectra for different bubble wall velocities. The upper left panel presents the results for $\tilde{v}_W$ = 0.4, the upper right panel for $\tilde{v}_W$ = 0.6, and the bottom panel for $\tilde{v}_W$ = 0.9. }
    \label{figgrav}
\end{figure}

In FIG. \ref{figgrav}, we present the GW spectra for the weak detonation (upper left), hybrid (upper right), and weak deflagration (bottom) modes, along with the sensitivity curves of LISA \cite{lisa}, Taiji \cite{taiji}, TianQin \cite{tianqin}, Big Bang Observer (BBO) \cite{bbodecigo}, DECi-hertz Interferometer GW Observatory (DECIGO) \cite{bbodecigo} and Ultimate-DECIGO (U-DECIGO) \cite{udecigo}. The peak spectrum value for BP1 is below $10^{-18}$ and is not displayed. For the other BPs, the GW spectra from the final step of PT are shown, as the durations of other steps are too short to produce significant signals. 
In the context of the hybrid mode, the peak spectrum value for BP4 at $T_{n2}$ falls within the sensitivity range of LISA detectors, which have peak frequencies around 0.001 Hz. 
The peak values of the GW spectrum for BP2, BP3, and BP5 at $T_{n2}$ fall within the sensitivity range of U-DECIGO, with amplitudes on the order of $\mathcal{O}(10^{-17})$ to $\mathcal{O}(10^{-15})$.

\section{Conclusion}\label{sec6}

We examined the multi-step PTs in the 2HDM+$\bm{a}$ considering the $\overline {{\rm{MS}}} $ scheme to renormalize the one-loop effective potential which effectively avoids the IR divergence associated with the Goldstone boson loop. After considering both theoretical and experimental constraints on this model, we identify four classes of evolution processes with multi-step PTs. In the first case, only the CP-even Higgs acquires a non-zero VEV and breaks electroweak symmetry while the CP symmetry is always conserved. In the remaining three cases, the pseudoscalar fields can obtain VEVs at different phases of the multi-step PTs, leading to the spontaneous breaking of the CP symmetry. As the temperature decreases, the observed vacuum at the present temperature is obtained via strong first-order electroweak PT, and the CP symmetry is recovered.
Finally, we compared the GW spectra generated in the four scenarios with the projected detection limits.
We found that the peaks of GW spectra from the final steps of the PT in BP2, BP3, BP4, and BP5 are expected to reach detection sensitivity when the bubble wall velocity is set to 0.6.

\section*{Acknowledgments}
We thank Yang Xiao for the helpful discussions. This work was supported by the National Natural Science Foundation
of China under grants,11975013, 12105248, and 12335005, by the project ZR2024MA001 and ZR2023MA038 supported by Shandong Provincial Natural Science Foundation, and Henan Postdoctoral Science Foundation HN2024003.
\newpage
\begin{appendix}
\section{The background-field-dependent masses and the thermal Debye masses}
We obtain the field-dependent masses of the scalars $h,H,A,\bm{a},H^ \pm$ and the Goldstone bosons ${G^ \pm },G$ by
\begin{equation}
	\begin{aligned}
		{\rm{ }}\hat m_{h,H,A,\bm{a},G}^2 &=& {\rm{eigenvalues}}\left( {\hat {\cal M}_N^2} \right),\\
		\hat m_{{G^ \pm },{H^ \pm }}^2 &=& {\rm{eigenvalues}}\left( {\hat {\cal M}_C^2} \right).
	\end{aligned}
\end{equation}
The $\hat {\cal M}_N^2$ denotes the mass matrix of the neutral Higgs, 
\begin{equation}
	\begin{aligned}
		&{\hat {\cal M}_{N11}^2 = \frac{{3{\lambda _1}}}{2}h_1^2 + \frac{{{\lambda _{345}}}}{2}h_2^2 + \frac{{{k_{345}}}}{2}h_3^2 + m_{11}^2 + \frac{{{\kappa _1}}}{2}h_4^2,{\rm{        }}}\\
		&{\hat {\cal M}_{N22}^2 = \frac{{3{\lambda _2}}}{2}h_2^2 + \frac{{{\lambda _2}}}{2}h_3^2 + \frac{{{\lambda _{345}}}}{2}h_1^2 + m_{22}^2 + \frac{{{\kappa _2}}}{2}h_4^2,{\rm{          }}}\\
		&\hat {\cal M}_{N33}^2 = \frac{{{\lambda _2}}}{2}h_2^2 + \frac{{3{\lambda _2}}}{2}h_3^2 + m_{22}^2 + \frac{{{k_{345}}}}{2}h_1^2 + \frac{{{\kappa _2}}}{2}h_4^2,{\rm{    }}\\
		&\hat {\cal M}_{N44}^2 = m_0^2 + \frac{{{\kappa _1}}}{2}h_1^2 + \frac{{{\kappa _2}}}{2}h_2^2 + \frac{{{\kappa _2}}}{2}h_3^2 + \frac{{{\kappa_S}}}{2}{h_4^2},{\rm{                 }}\\		
		&\hat {\cal M}_{N55}^2 = \frac{{{\lambda _1}}}{2}h_1^2 + m_{11}^2 + \frac{{{k_{345}}}}{2}h_2^2 + \frac{{{\lambda _{345}}}}{2}h_3^2 + \frac{{{\kappa _1}}}{2}{h_4^2},{\rm{   }}\\
		&{\hat {\cal M}_{N12}^2 = {\rm{ }}\hat {\cal M}_{N21}^2 = {\lambda _{345}}{h_1}{h_2} - m_{12}^2,{\rm{     }}}\\	
		&{\hat {\cal M}_{N13}^2 = {\rm{ }}\hat {\cal M}_{N31}^2 = {k_{345}}{h_1}{h_3}{\rm{ + }}{h_4}\mu, {\rm{                }}}\\	
		&\hat {\cal M}_{N14}^2 = {\rm{ }}\hat {\cal M}_{N41}^2 = {\kappa _1}{h_1}{h_4} + {h_3}\mu, {\rm{       }}\\	
		&\hat {\cal M}_{N15}^2 = {\rm{ }}\hat {\cal M}_{N51}^2 = {h_2}{h_3}{\lambda _5},{\rm{              }}\\				
		&\hat {\cal M}_{N23}^2 = {\rm{ }}\hat {\cal M}_{N32}^2 = {h_2}{h_3}{\lambda _2},{\rm{                }}\\
		&\hat {\cal M}_{N24}^2 = {\rm{ }}\hat {\cal M}_{N42}^2 = {\kappa _2}{h_2}{h_4},{\rm{                }}\\			
		&\hat {\cal M}_{N25}^2 = {\rm{ }}\hat {\cal M}_{N52}^2 = {h_1}{h_3}{\lambda _5} - {h_4}\mu, {\rm{               }}\\				
		&\hat {\cal M}_{N34}^2 = \hat {\cal M}_{N43}^2 = {\kappa _2}{h_3}{h_4} + {h_1}\mu, {\rm{               }}\\
		&\hat {\cal M}_{N35}^2 = \hat {\cal M}_{N53}^2 = {\lambda _5}{h_1}{h_2} - m_{12}^2,{\rm{         }}\\		
		&{\hat {\cal M}_{N45}^2 = \hat {\cal M}_{N54}^2 =  - {h_2}\mu. {\rm{                }}}
	\end{aligned}
\end{equation}
At zero temperature, due to $\langle h_3 \rangle=0$ and $\langle h_4 \rangle=0$, the $5\times 5$ matrix $\hat {\cal M}_N^2$ is reduced to
one $2\times 2$ matrix $\hat {\cal M}_P^2$ and one $3\times 3$ matrix $\hat {\cal M}_O^2$, with the former denoting the mass matrix of the CP-even scalar and the latter denoting the mass matrix of the CP-odd scalar, respectively. 
When solving Eqs.~(\ref{renormalization-1}) to (\ref{renormalization-4}), the background field also includes 
 $\phi_1^\pm$, $\phi_2^\pm$ and $\eta_1$.
 The $\hat {\cal M}_C^2$ denotes the mass matrix of the charged Higgs,
\begin{equation}
	\begin{aligned}
		&\hat {\cal M}_{C11}^2 = \frac{{{\lambda _1}}}{2}h_1^2 + m_{11}^2 + \frac{{{\lambda _3}}}{2}h_2^2 + \frac{{{\lambda _3}}}{2}h_3^2 + \frac{{{\kappa _1}}}{2}h_4^2,{\rm{    }}\\
		&\hat {\cal M}_{C22}^2 = \frac{{{\lambda _2}}}{2}h_2^2 + \frac{{{\lambda _2}}}{2}h_3^2 + m_{22}^2 + \frac{{{\lambda _3}}}{2}h_1^2 + \frac{{{\kappa _2}}}{2}h_4^2,{\rm{   }}\\
		&\hat {\cal M}_{C12}^2 = {\left( {\widehat {\cal M}_{C21}^2} \right)^\dag } = \frac{{\left( {{\lambda _4} + {\lambda _5}} \right)}}{2}{h_1}{h_2} - m_{12}^2 + i{h_4}\mu  + i\frac{{\left( {{\lambda _4} - {\lambda _5}} \right)}}{2}{h_1}{h_3}{\rm{     }}.
	\end{aligned}
\end{equation}
 The field-dependent masses of the gauge bosons ${W^ \pm },Z,\gamma $ are
\begin{equation}
	\begin{aligned}
		&\hat m_{{W^ \pm }}^2 = \frac{1}{4}{g^2}\left( {h_1^2 + h_2^2 + h_3^2} \right),\\
		&\hat m_Z^2 = \frac{1}{4}\left( {{g^2} + {g^{\prime 2}}} \right)\left( {h_1^2 + h_2^2 + h_3^2} \right),\\
		&\hat m_\gamma ^2 = 0.
	\end{aligned}
\end{equation}
Neglecting the contribution of lighter fermions, we focus on the field-dependent masses of the top quark,
\begin{equation}
\hat m_t^2 = \frac{1}{2}y_t^2\left[\left( h_1 c_\beta + h_2 s_\beta\right)^2 + h_3^2  s_\beta^2\right],
\end{equation}  
where ${y_t} = \sqrt 2 {m_t}/v$.

The thermal Debye masses $\bar M_i^2\left( {{h_1},{h_2},{h_3},{h_4},T} \right)$, where $i=h, H, A, \bm{a}, G, H^\pm, G^\pm$, are the eigenvalues of the full mass matrix, 
 \begin{equation}
 	\begin{aligned}
 		&\bar M_i^2\left( {{h_1},{h_2},{h_3},{h_4},T} \right) = {\rm{eigenvalues}}\left[ {\widehat {{\cal M}_X^2}\left( {{h_1},{h_2},{h_3},{h_4}} \right) + {\Pi _X}\left( T \right)} \right],\\
 		&{\Pi _{P11}} = \left[ {\frac{{9{g^2}}}{2} + \frac{{3{g^{\prime 2}}}}{2} + {6y_t^2}{c_\beta ^2} + 6{\lambda _1} + 4{\lambda _3} + 2{\lambda _4} + {\kappa _1}} \right]\frac{{{T^2}}}{{24}},\\
 		&{\Pi _{P22}} = \left[ {\frac{{9{g^2}}}{2} + \frac{{3{g^{\prime 2}}}}{2} + {6y_t^2}{s_\beta ^2} + 6{\lambda _2} + 4{\lambda _3} + 2{\lambda _4} + {\kappa _2}} \right]\frac{{{T^2}}}{{24}},\\
 		&{\Pi _{P33}} = {\Pi _{P22}},\\
 		&{\Pi _{P44}} = \left[ {4{\kappa _1} + 4{\kappa _2} + {\kappa _S}} \right]\frac{{{T^2}}}{{24}},\\
 		&{\Pi _{P55}} = {\Pi _{P11}},\\
 		&{\Pi _{C11}} = {\Pi _{P11}},\\
 		&{\Pi _{C22}} = {\Pi _{P22}}.
 	\end{aligned}
 \end{equation}
 The corrected thermal mass of the longitudinally polarized $W$ boson is
 \begin{equation}
 	\bar M_{W_L^ \pm }^2 = \frac{1}{4}{g^2}\left( {h_1^2 + h_2^2 + h_3^2} \right) + 2{g^2}{T^2}.
 \end{equation}
 The corrected thermal mass of the longitudinally polarized $Z$ and $\gamma$ boson
 \begin{equation}
 	\bar M_{{Z_L},{\gamma _L}}^2 = \frac{1}{8}\left( {{g^2} + {g^{\prime 2}}} \right)\left( {h_1^2 + h_2^2 + h_3^2} \right) + \left( {{g^2} + {g^{\prime 2}}} \right){T^2} \pm \Delta ,
 \end{equation}
 with
 \begin{equation}
 	{\Delta ^2} = \frac{1}{{64}}{\left( {{g^2} + {g^{\prime 2}}} \right)^2}{\left( {h_1^2 + h_2^2 + h_3^2 + 8{T^2}} \right)^2} - {g^2}{g^{\prime 2}}{T^2}\left( {h_1^2 + h_2^2 + h_3^2 + 4{T^2}} \right).
 \end{equation}

\end{appendix}

\end{document}